\documentclass[
    ,final            
  ]
  {aipproc}

\layoutstyle{8x11single}

\begin{document}

\title{A Transport Model for Nuclear Reactions Induced by Radioactive Beams}



\author{Bao-An Li}{address={Department of Chemistry and Physics, 
Arkansas State University, State University, AR 72467-0419, USA}}

\author{Lie-Wen Chen}{address={Institute of Theoretical Physics, 
Shanghai Jiao Tong University, Shanghai 200240, P.R. China}}

\author{Champak B. Das
\footnote{Present address: Variable Energy Cyclotron Center, 1/AF, Bidhannagar, Kolkata 700064, India}}
{address={Physics Department, McGill University, Montreal, Canada H3A 2T8}}
 
\author{Subal Das Gupta}{address={Physics Department, McGill University, Montreal, Canada H3A 2T8}}

\author{Charles Gale}{address={Physics Department, McGill University, Montreal, Canada H3A 2T8}}

\author{Che Ming Ko}{address={Cyclotron Institute and Physics Department, Texas A\& M University, 
College Statio, TX 77843, USA}}

\author{Gao-Chan Yong}{address={Institute of Modern Physics, 
Chinese Academy of Science, Lanzhou 730000, P.R. China}}

\author{Wei Zuo}{address={Institute of Modern Physics, 
Chinese Academy of Science, Lanzhou 730000, P.R. China}}

\begin{abstract}
Major ingredients of an isospin and momentum dependent transport model for
nuclear reactions induced by radioactive beams are outlined. Within the
IBUU04 version of this model we study several experimental probes of the
equation of state of neutron-rich matter, especially the density dependence
of the nuclear symmetry energy. Comparing with the recent experimental data
from NSCL/MSU on isospin diffusion, we found a nuclear symmetry energy of $%
E_{sym}(\rho )\approx 31.6(\rho /\rho_{0})^{1.05}$ at subnormal densities.
Predictions on several observables sensitive to the density dependence of
the symmetry energy at supranormal densities accessible at GSI and the
planned Rare Isotope Accelerator (RIA) are also made.
\end{abstract}

\maketitle

\section{Introduction}

\label{intro} The Equation of State (EOS) of isospin asymmetric nuclear
matter can be written within the well-known parabolic approximation, which
has been verified by all many-body theories, as 
\begin{equation}  \label{ieos}
E(\rho ,\delta )=E(\rho ,\delta =0)+E_{sym}(\rho )\delta ^{2}+\mathcal{O}%
(\delta^4),
\end{equation}
where $\delta\equiv(\rho_{n}-\rho _{p})/(\rho _{p}+\rho _{n})$ is the
isospin asymmetry and $E_{sym}(\rho)$ is the density-dependent nuclear
symmetry energy. The latter is very important for many interesting
astrophysical problems\cite{lat01}, the structure of radioactive nuclei\cite%
{brown,stone} and heavy-ion reactions\cite{ireview98,ibook01,dan02,ditoro}.
Unfortunately, the density dependence of symmetry energy $E_{sym}(\rho)$,
especially at supranormal densities, is still poorly known. Predictions
based on various many-body theories diverge widely at both low and high
densities. In fact, even the sign of the symmetry energy above $3\rho_0$
remains uncertain\cite{bom1}. Fortunately, heavy-ion reactions, especially
those induced by radioactive beams, provide a unique opportunity to pin down
the density dependence of nuclear symmetry energy in terrestrial
laboratories. Significant progress in determining the symmetry energy at
subnormal densities has been made recently both experimentally and
theoretically\cite{betty04,chen05}. High energy radioactive beams to be
available at GSI and RIA will allow us to determine the symmetry energy at
supranormal densities.

To extract information about the EOS of neutron-rich matter from nuclear
reactions induced by radioactive beams, one needs reliable theoretical
tools. Transport models are especially useful for this purpose. Especially
for central, energetic reactions at RIA and GSI, transport models are the
most useful tool for understanding the role of isospin degree of freedom in
the reaction dynamics and extract information about the EOS of neutron-rich
matter. Significant progresses have been made recently in improving
semi-classical transport models for nuclear reactions. While developing
practically implementable quantum transport theories is a long term goal,
applications of the semi-classical transport models have enabled us to learn
a great deal of interesting physics from heavy-ion reactions. In the
following we outline the major ingredients of an isospin and momentum
dependent transport model applicable for heavy-ion reactions induced by both
stable and radioactive beams\cite{lidas03}. This model has been found very
useful in understanding a number of new phenomena associated with the
isospin degree of freedom in heavy-ion reactions. Based on applications of
this model, we highlight here the most recent progress in determining the
symmetry energy at subnormal densities and present our predictions on
several most sensitive probes of the symmetry energy at supranormal
densities.

\begin{figure}[tbh]
\includegraphics{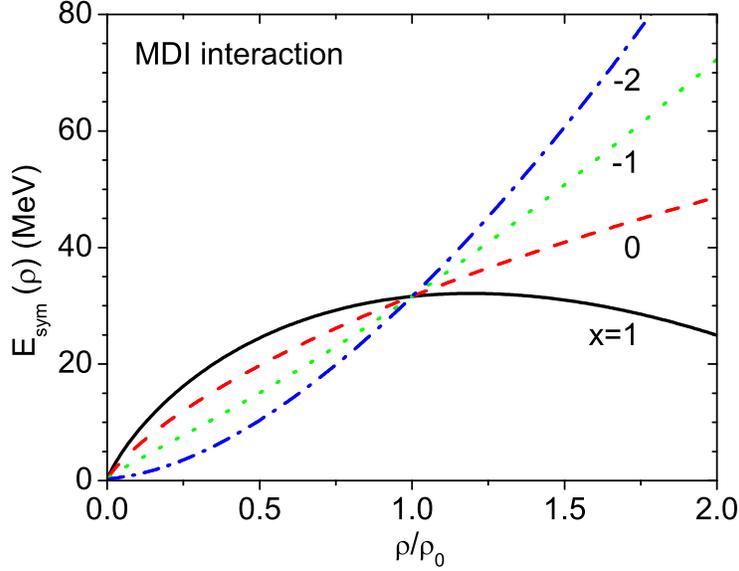} \vspace*{-1cm}
\caption{{\protect\small Density dependence of the symmetry energy for four }%
$x${\protect\small \ parameters.}}
\label{fig1}
\end{figure}

\section{IBUU04 version of the model}

Crucial to the extraction of critical information about the $E_{\mathrm{sym}%
}(\rho )$ is to compare experimental data with transport model calculations.
We outline here the major ingredients of the version IBUU04 of an isospin-
and momentum-dependent transport model for nuclear reactions induced by
radioactive beams\cite{lidas03}. The single nucleon potential is one of the
most important inputs to all transport models. Both the isovector (symmetry
potential) and isoscalar parts of this potential should be momentum
dependent due to the non-locality of strong interactions and the Pauli
exchange effects in many-fermion systems. In the IBUU04, we use a single
nucleon potential derived from the Hartree-Fock approximation using a
modified Gogny effective interaction (MDI)\cite{das03}, i.e., 
\begin{eqnarray}
U(\rho ,\delta ,\vec{p},\tau ,x) &=&A_{u}(x)\frac{\rho _{\tau ^{\prime }}}{%
\rho _{0}}+A_{l}(x)\frac{\rho _{\tau }}{\rho _{0}}  \nonumber  \label{mdi} \\
&+&B(\frac{\rho }{\rho _{0}})^{\sigma }(1-x\delta ^{2})-8\tau x\frac{B}{%
\sigma +1}\frac{\rho ^{\sigma -1}}{\rho _{0}^{\sigma }}\delta \rho _{\tau
^{\prime }}  \nonumber \\
&+&\frac{2C_{\tau ,\tau }}{\rho _{0}}\int d^{3}p^{\prime }\frac{f_{\tau }(%
\vec{r},\vec{p}^{\prime })}{1+(\vec{p}-\vec{p}^{\prime })^{2}/\Lambda ^{2}} 
\nonumber \\
&+&\frac{2C_{\tau ,\tau ^{\prime }}}{\rho _{0}}\int d^{3}p^{\prime }\frac{%
f_{\tau ^{\prime }}(\vec{r},\vec{p}^{\prime })}{1+(\vec{p}-\vec{p}^{\prime
})^{2}/\Lambda ^{2}}.
\end{eqnarray}%
In the above $\tau =1/2$ ($-1/2$) for neutrons (protons) and $\tau \neq \tau
^{\prime }$; $\sigma =4/3$; $f_{\tau }(\vec{r},\vec{p})$ is the phase space
distribution function at coordinate $\vec{r}$ and momentum $\vec{p}$. The
parameters $A_{u}(x),A_{l}(x),B,C_{\tau ,\tau },C_{\tau ,\tau ^{\prime }}$
and $\Lambda $ were obtained by fitting the momentum-dependence of the $%
U(\rho ,\delta ,\vec{p},\tau ,x)$ to that predicted by the Gogny
Hartree-Fock and/or the Brueckner-Hartree-Fock calculations, the saturation
properties of symmetric nuclear matter and the symmetry energy of 30 MeV at
normal nuclear matter density $\rho _{0}=0.16$ fm$^{-3}$\cite{das03}. The
incompressibility $K_{0}$ of symmetric nuclear matter at $\rho _{0}$ is set
to be 211 MeV. The parameters $A_{u}(x)$ and $A_{l}(x)$ depend on the $x$
parameter according to 
\begin{equation}
A_{u}(x)=-95.98-x\frac{2B}{\sigma +1},~~~~ A_{l}(x)=-120.57+x\frac{2B}{%
\sigma +1}.
\end{equation}
The parameter $x$ can be adjusted to mimic predictions on the $E_{sym}(\rho )
$ by microscopic and/or phenomenological many-body theories. The last two
terms contain the momentum-dependence of the single-particle potential. The
momentum dependence of the symmetry potential stems from the different
interaction strength parameters $C_{\tau ,\tau ^{\prime }}$ and $C_{\tau
,\tau }$ for a nucleon of isospin $\tau $ interacting, respectively, with
unlike and like nucleons in the background fields. More specifically, we use 
$C_{unlike}=-103.4$ MeV and $C_{like}=-11.7$ MeV. As an example, shown in
Fig.\ 1 is the density dependence of the symmetry energy for $x=-2$, $-1$, $0
$ and $1$. 
\begin{figure}[tbh]
\vspace*{-0.8cm} \includegraphics{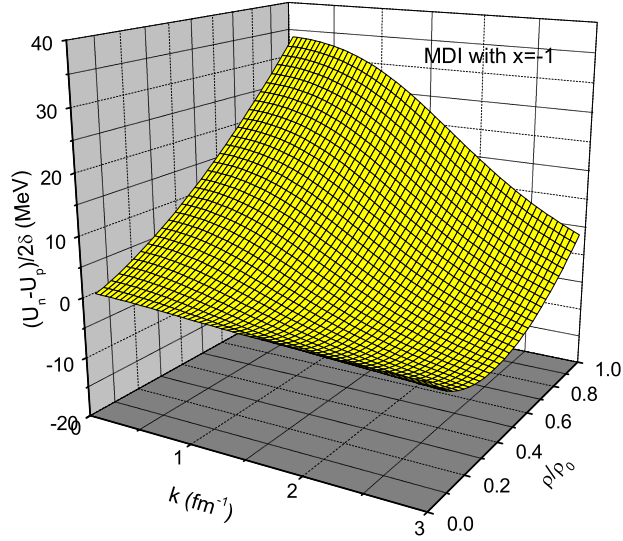} \vspace*{-1cm}
\caption{{\protect\small Symmetry potential as a function of momentum and
density for MDI interaction with }$x=-1${\protect\small .}}
\label{fig2}
\end{figure}

Systematic analyses of a large number of nucleon-nucleus and (p,n) charge
exchange scattering experiments at beam energies below about 100 MeV
indicate undoubtedly that the symmetry potential at $\rho _{0}$, i.e., the
Lane potential, decreases approximately linearly with increasing beam energy 
$E_{kin}$ according to $U_{Lane}=a-bE_{kin}$ where $a\simeq 22-34$ MeV and $%
b\simeq 0.1-0.2$\cite{data1,data2}. This provides a stringent constraint on
the symmetry potential. The potential in eq.\ref{mdi} meets this requirement
very well as seen in Fig.\ 2 where the symmetry potential $%
(U_{n}-U_{p})/2\delta $ as a function of momentum and density for the
parameter $x=-1$ is displayed.

One characteristic feature of the momentum dependence of the symmetry
potential is the different effective masses for neutrons and protons in
isospin asymmetric nuclear matter, i.e., 
\begin{equation}
\frac{m_{\tau }^{\ast }}{m_{\tau }}=\left\{ 1+\frac{m_{\tau }}{\hbar ^{2}k}%
\frac{dU_{\tau }}{dk}\right\} _{k=k_{\tau }^{F}}^{-1},  \label{mstar}
\end{equation}%
where $k_{\tau }^{F}$ is the nucleon Fermi wave number. With the potential
in eq. \ref{mdi}, since the momentum-dependent part of the nuclear potential
is independent of the parameter $x$, the nucleon effective masses are
independent of the $x$ parameter too. Shown in Fig. 3 are the nucleon
effective masses as a function of density (upper window) and isospin
asymmetry (lower window). It is seen that the neutron effective mass is
higher than the proton effective mass and the splitting between them
increases with both the density and isospin asymmetry of the medium\cite%
{lidas03}. We notice here that the momentum dependence of the symmetry
potential and the associated splitting of nucleon effective masses in
isospin asymmetric matter is still highly contoversial\cite{ditoro04,li04}.
The experimental determination of both the density and momentum dependence
of the symmetry potential is required. Please see also ref.\cite{ditoro05}
on this point. 
\begin{figure}[tbh]
\vspace*{-0.8cm} \includegraphics[height=0.45%
\textheight,angle=-90]{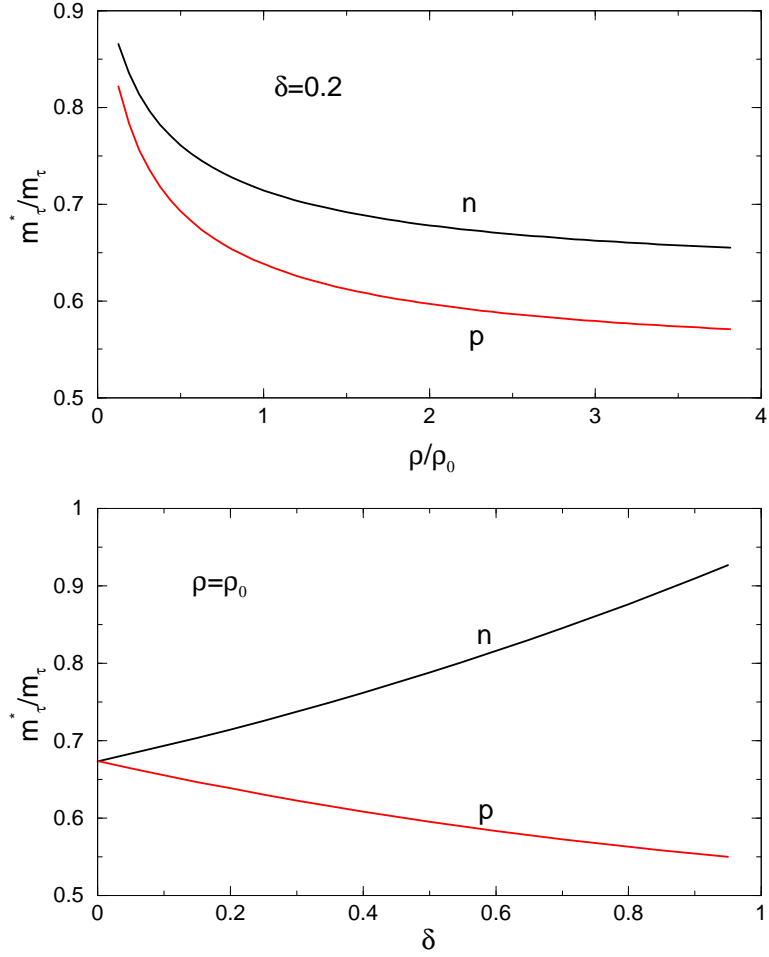} \vspace*{-1cm}
\caption{{\protect\small Nucleon effective masses in asymmetric matter as a
function of density (upper window) and isospin asymmetry (lower window).}}
\label{figure3}
\end{figure}

\begin{figure}[tbh]
\vspace*{-0.8cm} \includegraphics[height=0.45%
\textheight,angle=-90]{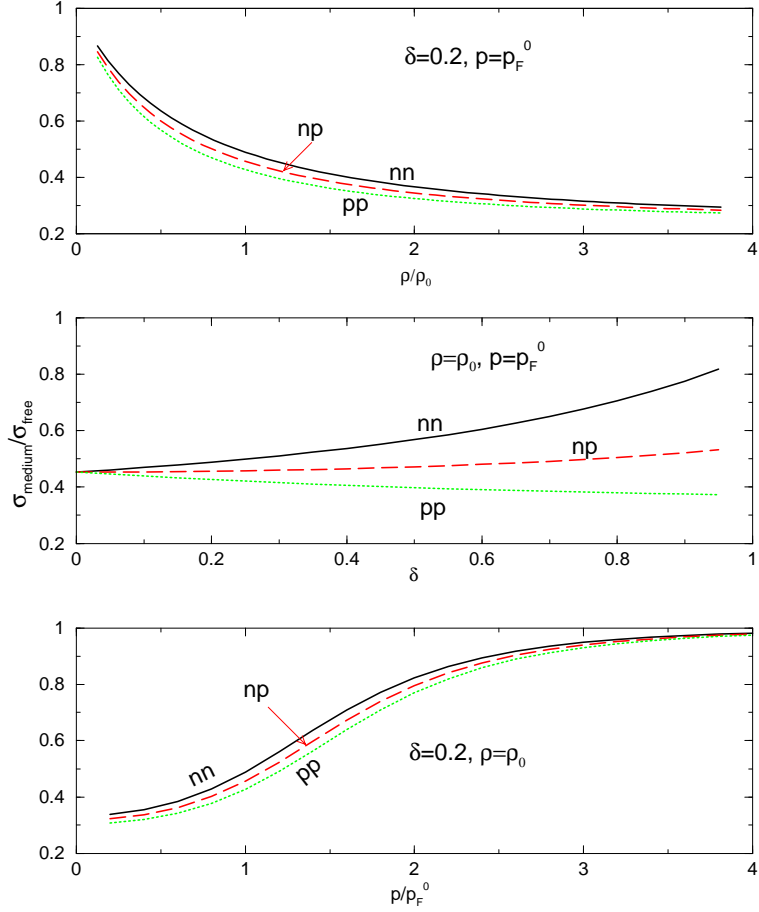} \vspace*{-1cm}
\caption{{\protect\small The reduction factor of the in-medium nucleon-nucleon cross sections with 
respect to the free space ones as a function of density (top), isospin asymmetry (middle) and 
momentum (bottom).}}
\label{figure4}
\end{figure}

Since both the incoming current in the initial state and the level density
of the final state in nucleon-nucleon (NN) scatterings depend on the
effective masses of colliding nucleons in medium, the in-medium
nucleon-nucleon cross sections are expected to be reduced by a factor 
\begin{equation}
\sigma _{NN}^{medium}/\sigma _{NN}=(\mu _{NN}^{\ast }/\mu _{NN})^{2}
\end{equation}%
where $\mu _{NN}$ and $\mu _{NN}^{\ast }$ are the reduced mass of the
colliding nucleon pairs in free-space and in medium, 
respectively\cite{sigma1,s2,s3}
effective masses is consistent with predictions based on more microscopic 
many-body theories\cite{fr}. 
Thus, because of the reduced in-medium nucleon effective masses and their
dependence on the density and isospin asymmetry of the medium, the in-medium
NN cross sections are not only reduced compared to their free-space values,
the nn and pp cross sections are split and the difference between them grows
in more asymmetric matter as shown in Fig. \ref{figure4}. The in-medium NN cross sections
are also independent of the parameter $x$. The isospin-dependence of the in-medium 
NN cross sections is expected to play an important role in nuclear reactions 
induced by neutron-rich nuclei\cite{li05}

Other details, such as the initialization of colliding nuclei in phase
space, Pauli blocking, etc can be found in our earlier publication\cite%
{ireview98,ibook01,lidas03}.

\section{Application of the model}

The model outlined above have many applications in heavy-ion reactions
induced by both stable and radioactive beams. In this section, we illustrate
several examples of studying the role of isospin degree of freedom in the
reaction dynamics and extracting the density dependence of nuclear symmetry
energy. Besides probes of symmetry energy at subnormal densities, several
probes sensitive to the high density behavior of the symmetry energy have
been also proposed largely based on transport model simulations.

\subsection{Probing the symmetry energy at subnormal densities with isospin
diffusion}

\label{diffusion} Tsang et al.\cite{betty04} recently studied the degree of
isospin diffusion in the reaction $^{124}$Sn + $^{112}$Sn by measuring\cite%
{gsi} 
\begin{equation}
R_{i}=\frac{2X_{^{124}{Sn}+^{112}{Sn}}-X_{^{124}{Sn}+^{124}{Sn}}-X_{^{112}{Sn%
}+^{112}{Sn}}}{X_{^{124}{Sn}+^{124}{Sn}}-X_{^{112}{Sn}+^{112}{Sn}}}
\label{Ri}
\end{equation}%
where $X$ is the average isospin asymmetry $\left\langle \delta
\right\rangle $ of the $^{124}$Sn-like residue. The data are indicated in
Fig. 5 together with the IBUU04 predictions about the time evolutions of $%
R_{i}$ and the average central densities calculated with $x=-1$ using both
the MDI and the soft Bertsch-Das Gupta-Kruse (SBKD) interactions. It is seen
that the isospin diffusion process occurs mainly from about $30$ fm/c to $80$
fm/c corresponding to an average central density from about $1.2\rho _{0}$
to $0.3\rho _{0}$. The experimental data from MSU are seen to be reproduced
nicely by the MDI interaction with $x=-1$, while the SBKD interaction with $%
x=-1$ leads to a significantly lower $R_{i}$ value\cite{chen05}. 
\begin{figure}[th]
\includegraphics{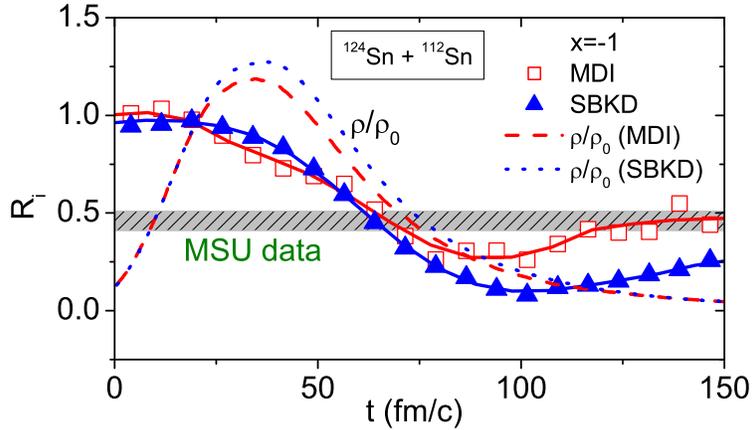}
\caption{{\protect\small The degree of isospin diffusion as a function of
time with the MDI and SBKD interactions. The corresponding evolutions of
central density are also shown.}}
\label{figure5}
\end{figure}

Effects of the symmetry energy on isospin diffusion were also studied by
varying the parameter $x$\cite{chen05}. Only with the parameter $x=-1$ the
data can be well reproduced. The corresponding symmetry energy can be
parameterized as $E_{sym}(\rho )\approx 31.6(\rho /\rho _{0})^{1.05}$. In
the present study on isospin diffusion, only the free-space NN cross
sections are used and thus effects completely due to the different density
dependence of symmetry energy are investigated. As the next step we are
currently investigating effects of the in-medium NN cross sections on the
isospin diffusion.

\subsection{Isospin asymmetry of dense matter formed in high energy heavy-ion reactions}

What are the maximum baryon density and isospin asymmetry that can be
achieved in central heavy-ion collisions at the highest beam energy expected
at RIA? This is an interesting question relevant to the study of the EOS of
asymmetric nuclear matter. To answer this question we show in Fig.\ 6 the
central baryon density (upper window) and the average $(n/p)_{\rho\geq
\rho_0}$ ratio (lower window) of all regions with baryon densities higher
than $\rho_0$ in the reaction of $^{132}Sn+^{124}Sn$ at a beam energy of 400
MeV/nucleon and an impact parameter of 1 fm. It is seen that the maximum
baryon density is about 2 times normal nuclear matter density. Moreover, the
compression is rather insensitive to the symmetry energy because the latter
is relatively small compared to the EOS of symmetric matter around this
density. The high density phase lasts for about 15 fm/c from 5 to 20 fm/c
for this reaction. It is interesting to see that the isospin asymmetry of
the high density region is quite sensitive to the symmetry energy. The soft
(e.g., $x=1$) symmetry energy leads to a significantly higher value of $%
(n/p)_{\rho\geq \rho_0}$ than the stiff one (e.g., $x=-2$). This is
consistent with the well-known isospin fractionation phenomenon. Because of
the $E_{sym}(\rho)\delta^2$ term in the EOS of asymmetric nuclear matter, it
is energetically more favorable to have a higher isospin asymmetry $\delta$
in the high density region with a softer symmetry energy functional $%
E_{sym}(\rho)$. In the supranormal density region, as shown in Fig.\ 1, the
symmetry energy changes from being soft to stiff when the parameter $x$
varies from 1 to -2. Thus the value of $(n/p)_{\rho\ge \rho_0}$ becomes
lower as the parameter $x$ changes from 1 to -2. It is worth mentioning that
the initial value of the quantity $(n/p)_{\rho\ge \rho_0}$ is about 1.4
which is less than the average n/p ratio of 1.56 of the reaction system.
This is because of the neutron-skins of the colliding nuclei, especially
that of the projectile $^{132}Sn$. In the neutron-rich nuclei, the n/p ratio
on the low-density surface is much higher than that in their interior. It is
clearly seen that the dense region can become either neutron-richer or
neutron-poorer with respect to the initial state depending on the symmetry
energy functional $E_{sym}(\rho)$ used.

\begin{figure}[th]
\includegraphics[height=0.3\textheight]{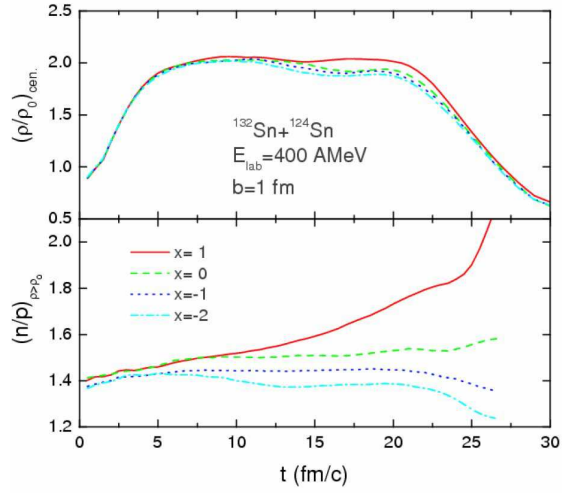}\vspace*{-1cm}
\caption{{\protect\small Central baryon density (upper window) and isospin
asymmetry (lower window) of high density region for the reaction of $%
^{132}Sn+^{124}Sn$ at a beam energy of 400 MeV/nucleon and an impact
parameter of 1 fm.}}
\label{figure6a}
\end{figure}

\subsection{Pions yields and $\protect\pi ^{-}/\protect\pi ^{+}$ ratio as a
probe of the symmetry energy at supranormal densities}

At the highest beam energy at RIA, pion production is significant. Pions may
thus carry interesting information about the EOS of dense neutron-rich matter%
\cite{lipi,gaopi}. Shown in Fig.7 are the $\pi ^{-}$ and $\pi ^{+}$ yields
as a function of the $x$ parameter. 
\begin{figure}[th]
\includegraphics[height=0.3\textheight]{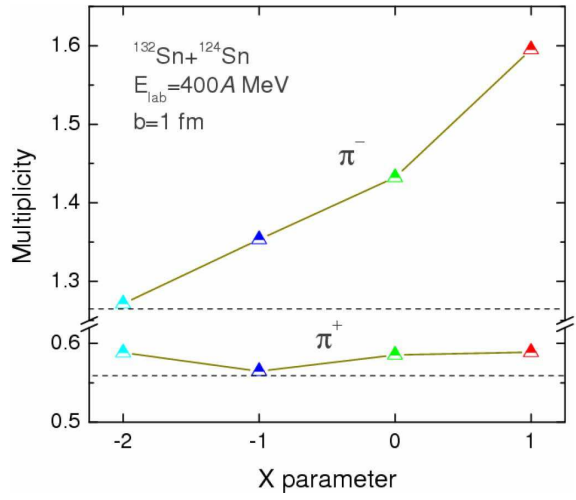}\vspace*{-1cm}
\caption{{\protect\small The $\protect\pi ^{-}$ and $\protect\pi ^{+}$
yields as functions of the $x$ parameter.}}
\label{figure7}
\end{figure}
It is interesting to see that the $\pi ^{-}$ multiplicity depends more
sensitively on the symmetry energy. The $\pi ^{-}$ multiplicity increases by
about 20\% while the $\pi ^{+}$ multiplicity remains about the same when the 
$x$ parameter is changed from -2 to 1. The multiplicity of $\pi ^{-}$ is
about 2 to 3 times that of $\pi ^{+}$. This is because the $\pi ^{-}$ mesons
are mostly produced from neutron-neutron collisions. Moreover, with the
softer symmetry energy the high density region is more neutron-rich due to
isospin fractionation\cite{gaopi}. The $\pi ^{-}$ mesons are thus more
sensitive to the isospin asymmetry of the reaction system and the symmetry
energy. However, one should notice that it is well known that pion yields
are also sensitive to the symmetric part of the nuclear EOS. It is thus hard
to get reliable information about the symmetry energy from $\pi ^{-}$ yields
alone. Fortunately, the $\pi ^{-}/\pi ^{+}$ ratio is a better probe since
statistically this ratio is only sensitive to the difference in the chemical
potentials for neutrons and protons\cite{bertsch}. This expectation is well
demonstrated in Fig.\ 8. It is seen that the pion ratio is quite sensitive
to the symmetry energy, especially at low transverse momenta. Thus, it is
promising that the high density behavior of nuclear symmetry energy $%
E_{sym}(\rho )$ can be probed using the $\pi ^{-}/\pi ^{+}$ ratio. 
\begin{figure}[th]
\includegraphics[height=0.3\textheight]{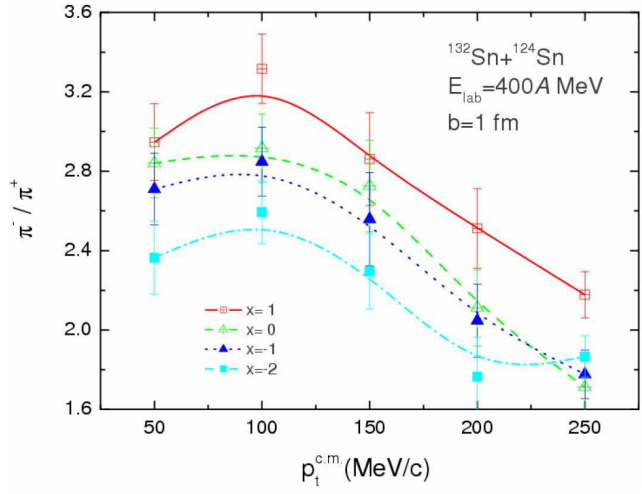}\vspace*{-1cm}
\caption{{\protect\small The $\protect\pi ^{-}/\protect\pi ^{+}$ ratio as a
function of transverse momentum.}}
\label{figure8}
\end{figure}

\subsection{Isospin fractionation and n-p differential flow at RIA and GSI}

\label{RIA2}

The degree of isospin equilibration or translucency can be measured by the
rapidity distribution of nucleon isospin asymmetry $\delta _{free}\equiv
(N_{n}-N_{p})/(N_{n}+N_{p})$ where $N_{n}$ ($N_{p}$) is the multiplicity of
free neutrons (protons)\cite{li04a}. Although it might be difficult to
measure directly $\delta _{free}$ because it requires the detection of
neutrons, similar information can be extracted from ratios of light
clusters, such as, $^{3}H/^{3}He$, as demonstrated recently within a
coalescence model\cite{chen03b,chen04a}. Shown in Fig.\ 9 are the rapidity
distributions of $\delta _{free}$ with (upper window) and without (lower
window) the Coulomb potential. It is interesting to see that the $\delta
_{free}$ at midrapidity is particularly sensitive to the symmetry energy. As
the parameter $x$ increases from $-2$ to $1$ the $\delta _{free}$ at
midrapidity decreases by about a factor of 2. Moreover, the forward-backward
asymmetric rapidity distributions of $\delta _{free}$ with all four $x$
parameters indicates the apparent nuclear translucency during the reaction%
\cite{ligao}. 
\begin{figure}[th]
\includegraphics[height=0.45\textheight]{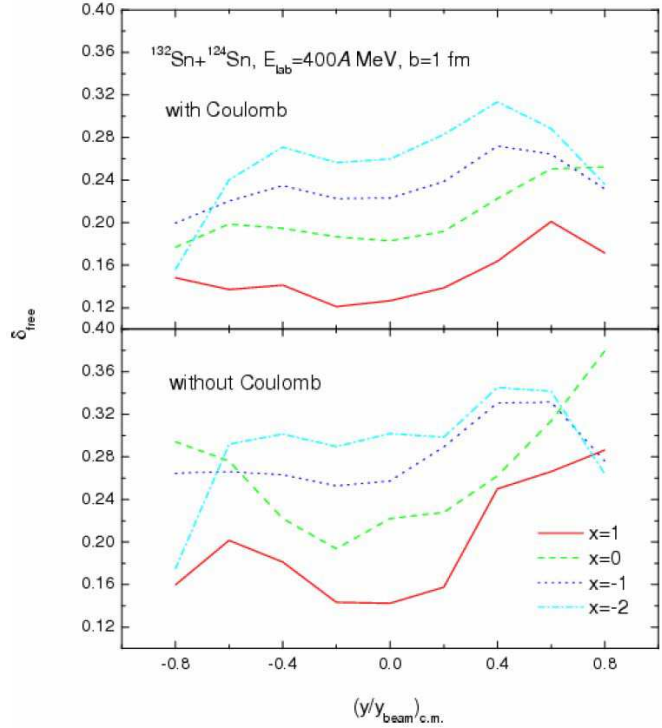}\vspace*{-1cm}
\caption{{\protect\small Isospin asymmetry of free nucleons with and without
the Coulomb force.}}
\label{figure9}
\end{figure}

Another observable that is sensitive to the high density behavior of
symmetry energy is the neutron-proton differential flow\cite{li00} 
\begin{equation}
F_{n-p}^{x}(y)\equiv \sum_{i=1}^{N(y)}(p_{i}^{x}w_{i})/N(y),
\end{equation}%
where $w_{i}=1(-1)$ for neutrons (protons) and $N(y)$ is the total number of
free nucleons at rapidity $y$. The differential flow combines constructively
effects of the symmetry potential on the isospin fractionation and the
collective flow. It has the advantage of maximizing the effects of the
symmetry potential while minimizing those of the isoscalar potential. Shown
in Fig. 10 is the n-p differential flow for the reaction of $%
^{132}Sn+^{124}Sn$ at a beam energy of 400 MeV/nucleon and an impact
parameter of 5 fm. Effects of the symmetry energy are clearly revealed by
changing the parameter $x$. 
\begin{figure}[th]
\includegraphics[height=0.3\textheight]{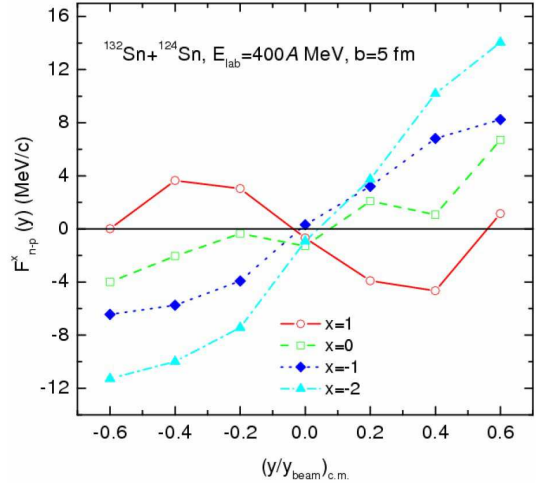}\vspace*{-1cm}
\caption{{\protect\small Neutron-proton differential flow at RIA and GSI
energies}}
\label{figure10}
\end{figure}

\section{Conclusions}
Transport models are powerful tools for investigating especially central 
heavy-ion reactions induced by neutron-rich nuclei. Applications of 
these models will help us understand 
the isospin dependence of in-medium nuclear effective interactions. Comparing with
the experimental data, we can extract the isospin dependence of thermal, mechanical and
transport properties of asymmetric nuclear matter playing important roles in nuclei, 
neutron stras and supernovae. Currently, the most important issue is the density
dependence of the nuclear symmetry energy. The latter is very important for both
nuclear physics and astrophysics. Significant progress has been made
recently by the heavy-ion community in determining the density dependence of
the nuclear symmetry energy. Based on transport model calculations, a number
of sensitive probes of the symmetry energy have been found. The momentum
dependence in both the isoscalar and isovector parts of the nuclear
potential was found to play an important role in extracting accurately the
density dependence of the symmetry energy. Comparing with recent
experimental data on isospin diffusion from NSCL/MSU, we have extracted a
symmetry energy of $E_{sym}(\rho )\approx 31.6(\rho /\rho _{0})^{1.05}$ at
subnormal densities. It would be interesting to compare this conclusion with
those extracted from studying other observables. More experimental data
including neutrons with neutron-rich beams in a broad energy range are
needed. Looking forward to experiments at RIA and GSI with high energy
radioactive beams, we hope to pin down the symmetry energy at supranormal
densities in the near future. Theoretically, the development of a practically
implementable quantum transport theory for nuclear reactions induced by
rdioactive beams remains a big challenge.

This work was supported in part by the US National Science Foundation of the
under grant No. PHY 0098805, PHYS-0243571 and PHYS0354572, Welch Foundation
grant No. A-1358, and the NASA-Arkansas Space Grants Consortium award
ASU15154. C.B. Das, S. Das Gupta and C. Gale were supported in part by the
Natural Sciences and Engineering Research Council of Canada, and the Fonds
Nature et Technologies of Quebec. L.W. Chen was supported by the National
Natural Science Foundation of China grant No. 10105008. The work of G.C.
Yong and W. Zuo was supported in part by the Chinese Academy of Science
Knowledge Innovation Project (KECK2-SW-N02), Major State Basic Research
Development Program (G2000077400), the National Natural Science Foundation
of China (10235030) and the Important Pare-Research Project (2002CAB00200)
of the Chinese Ministry of Science and Technology.

\vfill\eject

\end{document}